\documentclass[useAMS,usenatbib]{mn2e}
\usepackage{epsfig}

\title[]{ 
The statistics of accelerations seen in radial velocity searches for planets}
\author[]{Alice C. Quillen \\
{Department of Physics and Astronomy, University of Rochester, Rochester, NY 14627} \\
}
\begin{document}
\label{firstpage}
\maketitle

\begin{abstract}
Radial velocity searches for extrasolar planets have discovered 
stars undergoing slow accelerations. 
The accelerations are likely due to one or more gas giant planets 
or brown dwarfs orbiting with period longer than the total time span of
observations. The stellar acceleration is proportional 
to the mass of the companion divided
by the square of its radius from the star.
In this paper we predict the distribution 
of accelerations using a Monte Carlo method and assuming 
a semi-major axis and mass distribution for the companions. 
Radial velocity surveys suggest that  $20$ and  10\% of stars surveyed
exhibit accelerations above $\sim$ 10 and 25~m s$^{-1}$ yr$^{-1}$, respectively.
We find that an extrapolation of the size and period distribution
found by radial velocity surveys within $\sim 5$ AU or companion imaging 
surveys predicts too few systems with detectable accelerations.
The fractions of stars with
accelerations above these values can be matched
if every star hosts a gas giant planet or brown
dwarf with semi-major axes between 4 and 20 AU or if a significant fraction
($\sim$ 70\%) of stars host 3 or more giant planets in this range.
If brown dwarfs companions are not discovered in imaging
surveys then we would infer that there is a large population of 
gas giant planets present with
semi-major axes in the 5--30 AU range.

\end{abstract}

\section{Introduction}

Radial velocity searches for 
extrasolar planets often find linear
trends in the observed velocity of a star 
\citep{butler06,wittenmyer06,wright07,patel07,marcy08}.  
These correspond to velocity variations or constant
accelerations of the star by massive objects
in the system. As the time span of observations
increases these accelerations begin to vary (e.g., \citealt{wright07,patel07}), 
then finally a substantial fraction of the
an orbital period becomes recognizable
as the entire Keplerian velocity curve is recognized. 
Trends or accelerations have been
seen in radial velocity surveys \citep{butler06} and used to place
limits on mass and radii of long period
planets residing in the system \citep{wittenmyer06,wittenmyer07,wright07}.
Planets are not considered detected by 
radial velocity or transit surveys until at least
a full orbital period is seen \citep{otoole08}.  This is justifiable 
as it is difficult, perhaps impossible when noise is present,
to constrain simultaneously planet mass, semi-major axis and orbital
eccentricity from radial velocity 
observations alone over only a fraction of a period \citep{brown04,wright07},
but see the study by \citet{patel07}.
Once an entire period is observed,
the period of the orbit sets the planet's semi-major axis.
As the amplitude of the radial velocity perturbation
depends on both the mass and period, once
the period is known, the mass of the planet can be estimated.
Because only poor constraints on both planet and
semi-major axis can be made from constant or slowly changing
accelerations,
radial velocity surveys place only poor constraints 
on exoplanets at large semi-major axis
where the orbital periods are long.

In the next decade we have the prospect of
both more sensitive radial velocity surveys (nearing
1m/s in precision) and multi-object radial velocity surveys.
These surveys will increasingly find stars with measured accelerations
that don't correspond to complete orbital periods.
In this paper we consider the distribution of these trends
and how they can be used to constrain the mass
and semi-major axis distribution of
massive objects in outer exo-planetary systems even though
the exact masses and semi-major axes are not directly measured. 
A similar approach can be adopted for astrometric surveys
for long-period planets \citep{olling07}.
By outer we mean above 5~AU.  The orbital period at
5~AU around a solar mass star is about 11 years, corresponding
to the lifespan of a decade length 
multi-object radial velocity survey.
This semi-major axis is within the limit of most current
direct planet detection methods.  However the statistics of objects
inferred from the statistics of radial velocity trends
will be complimentary to and motivate direct detection surveys 
searching for brown dwarfs and planets at wider separations
(e.g., \citealt{metchev08,kraus08})
and transit surveys that primarily 
find planets within an AU of their host star.

\section{The acceleration of a star caused by a low mass companion}

We first consider the acceleration of a star
caused by  a single planet or brown dwarf in orbit.
In the center of mass frame of 2 gravitationally bound bodies with
masses $m_1$ and $m_2$ at positions $\vec{x_1}$ and $\vec{x_2}$,
the position of the more massive body is
\begin{equation}
\vec{x}_1 = {m_2 \over M} (\vec{x}_1 - \vec{x}_2)  = \mu \vec{r}
\end{equation}
where $M= m_1 + m_2$ is the sum of the masses of the two bodies,
the mass ratio $\mu =   m_2 /M$ and the vector between
the two bodies $\vec{r} = \vec{x}_1 - \vec{x}_2$.
The acceleration of the more massive body or host star
\begin{equation}
\ddot{\vec{x}}_1 =  \mu \ddot{\vec{r}}.
\end{equation}
For the two body Kepler problem 
\begin{equation}
\ddot{\vec{r}} = -{GM \hat{r} \over r^2}
\end{equation}
where $G$ is the gravitational constant
and $ \hat{r}$ is the unit vector between the two masses.
Combining the previous two equations we find that
the acceleration of the central star
\begin{equation}
\ddot{\vec{x}}_1 =-{\mu GM \hat{r} \over r^2}. 
\end{equation}
The acceleration toward the viewer measured
from changes in the radial velocity
is $a_r = \ddot{\vec{x}}_1 \cdot \hat{e}_{los}$
where $\hat{e}_{los}$ is the unit vector pointing toward the observer.
In physical units
\begin{equation}
a_r
 = 187 {\rm m~ s^{-1}~yr^{-1} }
     \left({r \over {\rm AU}}\right)^{-2}  
     \left({\mu \over 10^{-3}}\right)
     \left({M \over M_\odot}\right)
     \hat{r} \cdot \hat{e}_{los}.
\label{eqn:a_r}
\end{equation}
We note that the magnitude of the acceleration
depends on $\mu M /r^2 = M_p/r^2$, where $M_p$ is the mass
of the planet and is independent of the
stellar mass.  We expect the distribution
of accelerations to depend  on the ratio of the companion
mass divided by the square its semi-major axis. 
This dependence is reflected in the upper limit plots 
(mass vs period) by \citet{wittenmyer06}   (see their Figure 4)
derived from trends (accelerations) seen with the McDonald Observatory
planet search program.
For comparison the amplitude of the radial velocity
variation during an entire orbit is proportional to $\mu\sqrt{M/a}$
(e.g., equation 1 \citealt{cumming08}).
We expect the statistics of accelerations
to differ from those of radial velocity amplitudes.

The distribution of extrasolar planets discovered
in radial velocity surveys is consistent
with $dN/d \log a \propto a^{0.4}$ \citep{marcy08}
implying that there is increasing numbers of giant planets
at larger semi-major axes.
For a semi-major axis greater than 10~AU,
where an entire orbital period would not be detected
in a moderate time span radial velocity survey,
equation \ref{eqn:a_r}
predicts an acceleration of 
$ 1.9 {\rm m~ s^{-1}~yr^{-1} } (M_p/M_J)$ which
is near the detection limit of radial velocity
surveys.   Here $M_J$ is the mass
of Jupiter. However companion masses in the brown
dwarf range (between 13 and 80~$M_J$, \citealt{burrows97}) 
at semi-major
axes between 10 and 100~AU can cause
accelerations detectable in radial velocity surveys.
Hence accelerations measured from
radial velocity surveys are sensitive to
massive planets and brown dwarfs  in the outer regions
of extrasolar systems.  
This range of masses  contains brown dwarfs
and so the region identified as the brown dwarf desert 
where there is a deficit of known objects \citep{grether06}.
Previous studies have recognized that incomplete
orbits detected in radial velocity studies are
most likely caused by companion brown dwarfs \citep{patel07}.

\citet{cumming08} found that $\sim 23\%$ of the total
stars surveyed showed a significant slope or velocity
variations (corresponding to accelerations) across 8 years
of observations.
The mean error of their survey was less than 3 m/s 
so the velocity variations detected would be similar to
accelerations (along line of sight) are above few m~s$^{-1}$~yr$^{-1}$.
With less precision \citet{nidever02} found that 12\% of the stars
surveyed varied by more than 
above 100 m~s$^{-1}$ across 4 years, 
corresponding to accelerations above 25 m~s$^{-1}$~yr$^{-1}$.
A period of 4 years corresponds to a semi-major axis of 2.5~AU.
If we restrict the semi-major axis to greater than 3~AU
and use Equation \ref{eqn:a_r}, the level of accelerations seen
by \citet{nidever02}
requires planet masses above 1~$M_J$ if they are found at semi-major
axes above 3~AU.  
We have conservatively chosen a semi-major axis corresponding
to a period only slightly longer than the survey length.  
As the acceleration switches sign every half period
it likely the objects responsible for these accelerations
are at semi-major axis with periods larger than twice 
the survey length.
For the longer Keck planet search  8 years corresponds to a 
semi-major axis of 4~AU.   If we restrict the companion
semi-major axis to 6~AU then again a planet mass above 1~$M_J$
is required to account for the observed velocity variations.
The large fractions of objects exhibiting detected accelerations
suggests that there is significant population of massive objects in
the outskirts of exo-planetary systems.  This is consistent
with the extrapolation to larger semi-major
axes discussed by \citet{cumming08} who estimated the number
of systems with planets
based on size distribution of inner planets exhibiting full periods
during the span of observation.

\section{Monte Carlo simulations}

To compare the number of objects with accelerations
to those predicted from the size and period distributions
we use a Monte Carlo method (similar to the study by \citealt{brown04}).
We compute the line of sight component of the acceleration
$a_r$, based on a distribution of systems with planets. 
We first consider the distribution of accelerations that
would be seen from a very simple population.
We consider a group of systems, each with one
massive planet or brown dwarf companion 
at random positions in their orbits and 
at random inclinations.
Mean anomalies and arguments of perihelia 
are randomly chosen from a flat probability distribution.
Eccentricities are chosen from a flat eccentricity distribution
with minimum and maximum eccentricities $e_{min}$ and $e_{max}$.
Orbital inclinations are chosen by generating $\cos i$ with a flat
probability distribution between -1 and 1 (\citealt{brown04}, equation 17).
After randomly generating orbital elements the Cartesian
coordinates of each system are computed using Kepler's equation.
Then the coordinates of the system are projected into viewer
coordinates using the randomly selected system inclination.
Equation \ref{eqn:a_r} is used to predict the line of sight
component of the acceleration.
Acceleration distributions are computed using $10^5$ sample star systems.

Figure 1 shows the distribution of the radial or line
of sight component of
acceleration $a_r$ computed
for a population of systems with 10~$M_J$ 
bodies at 5 and 10~AU chosen with flat eccentricity distributions.
For each semi-major axis, we 
show eccentricity distributions that consist
of objects in circular orbits, range from 0 to
0.5 or 0 to 0.99.  As planets on high eccentricity orbits have
higher accelerations when they approach the star, the acceleration
distributions extend to higher velocities when the eccentricity
distribution is wider. 

\begin{figure}
\includegraphics[angle=0,width=3.5in]{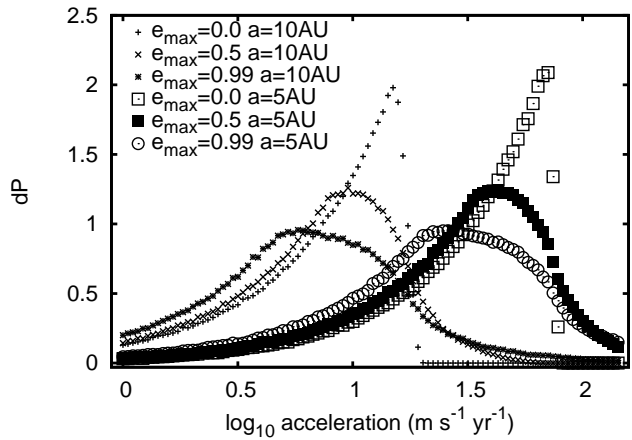}
\caption{
Probability distributions for the radial component
of acceleration  for a population of systems each with a
10~$M_J$ planets at
a semi-major axis of 5~AU (large points) and 10~AU (smaller points).
The orbital inclinations, mean motions and arguments
of perihelia are randomly chosen. 
Objects are drawn from 3 eccentricity distributions for each planet 
semi-major axis.  The first distribution is for circular
orbits, the second with a flat distribution ranging from eccentricity
$e=0$ to 0.5  and the third randing from $e=0$ to 0.99.
Higher eccentricities allow closer approaches and so
higher accelerations.   A lower mass planet would caused similar
curves but shifted to the left because the accelerations would be lower.
\label{fig:a}
}
\end{figure}

For these simple systems we ask what is the fraction
of objects that would be observed with accelerations
greater than a particular value, 5, 10 or 25 m~s$^{-1}$~yr$^{-1}$,
$f_5$, $f_{10}$ and $f_{25}$.
These fractions mimic the fraction of objects with trends reported
by \citet{cumming08,nidever02}.
These fractions are listed in Table \ref{tab:single} and
are measured from the same simulations
that made the acceleration distributions
shown in Figure \ref{fig:a}.

The fraction of stars, $f_5$, $f_{10}$ and $f_{25}$, with accelerations
above 5, 10 or 25 m~s$^{-1}$~yr$^{-1}$ is not high. 
The fraction $f_{25}$ does not approach the 12\%
based on  the survey by \citet{nidever02} 
except for the systems with masses above 5~$M_J$ mass planets at small
radii at or below 5 AU.
The fractions $f_5$ or $f_{10}$ do not approach 
the 23\% estimated (based on the report by \citealt{cumming08}) 
unless masses are above 5~$M_J$.
The fractions shown in Table \ref{tab:single}
were computed assuming that every system contains
a single massive planet.  The fraction of stars
showing accelerations is necessarily lower than
the fraction of systems with planets (here
assumed to be all systems) as some of
the systems are viewed face on and systems are
viewed at different times in their orbit.
The larger fraction of stars with accelerations
compared to those predicted with this
simple model suggests that
one massive object is required at 5-20AU
in almost all systems or more than one massive object
is required in a significant fraction of systems. 

\subsection{Accelerations predicted for a distribution
of companion masses and periods}

We now consider Monte Carlo simulations for systems
with planets and brown dwarfs of different masses
and semi-major axes.
The semi-major axis and planet mass distribution is assumed to
be power law in form and is drawn from a specified range
with limits $a_{min}$, $a_{max}$ and $M_{p,min}$ and $M_{p,max}$.  
This type of distribution function was
used to fit the radial velocity planet search measurements
at the Lick and Keck observatories and the Anglo Australian
Telescope by \citet{marcy08,cumming08}.
We assume that
the probability of a planet with semi-major axis $a$ and 
planet mass $M_p$
\begin{equation}
dP = C M_p^{\alpha} a^{\beta}~dM_p~da~de.
\label{eqn:dP}
\end{equation}
\citep{marcy08} find 
that $\alpha \approx -1.1$ and $\beta \approx -0.6$.
The exponent $\beta = -0.6$ is consistent with 
$dN/d\log a \propto a^{0.4}$.
\citet{marcy08} find a broad eccentricity distribution exterior to
$a=0.1$~AU with maximum eccentricity $e=0.93$.
The coefficient $C$ is such that $\sim 10.5\%$ have planets
with mass in range 0.3 to 10 MJ and orbital period in range
of 2 to 2000 days \citep{cumming08}.
\citet{cumming08} computed the distribution for $M_p \sin i$ instead
of $M_p$, however only
a small correction is needed as
$\langle \sin i \rangle = \pi/4 = 0.785$ (\citealt{cumming08}, section 3.0).

\begin{figure}
\includegraphics[angle=0,width=3.5in]{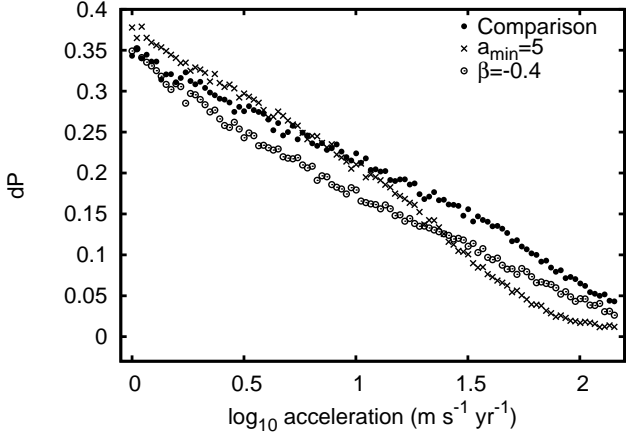}
\caption{
Probability distributions for the radial component
of acceleration  for a population of planets
of a given mass but for different semi-major axis
distributions.  The comparison model, (solid circles)
has 10~$M_J$ mass planets chosen
with a semi-major axis distribution $dN \propto a^{\beta}$ 
with $\beta=-0.6$.
The probability for the comparison model is computed assuming that all systems
have 10~$M_J$ planets with this distribution and with
semi-major axes in the range
between $a_{min} = 3$~AU and $a_{max}= 100$~AU
and eccentricities in the range 0 to 0.99.   
The probability distribution
is not sensitive to $a_{max}$.  Also shown
are distributions for a higher minimum
semi-major axis $a_{min} = 5$~AU and shallower exponent $\beta=-0.4$
with all other parameters the same as the comparison model. 
The probability distribution is only weakly sensitive to the inner
radius of the distribution and the exponent of the underlying 
semi-major axis distribution.
Power law distributions in the line of sight component 
of acceleration are predicted.
\label{fig:b}
}
\end{figure}

We now look at the distribution of accelerations for
a population of a fixed companion mass but with a distribution
of semi-major axes.
We consider randomly oriented systems populated by 10~$M_J$ planets with
semi-major axis chosen
from a probability distribution $dP \propto a^{\beta}$,
a flat eccentricity distribution ranging from $e=0$ to 0.99
and minimum and maximum semi-major axes between $a_{min}$
and $a_{max}$.
Figure \ref{fig:b} shows that the probability distribution
of the radial acceleration component
is only weakly dependent on the exponent of
the semi-major axis distribution and on the inner radius of
the distribution.  These curves would move to the left
for a population of lower mass planets as they
would cause lower accelerations. 

\begin{figure}
\includegraphics[angle=0,width=3.5in]{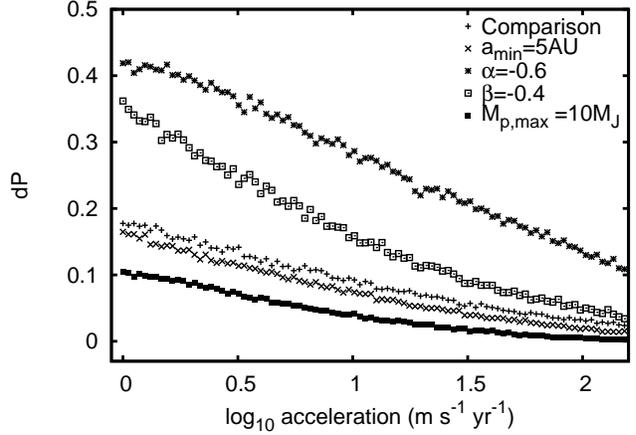}
\caption{
Normalized probability distributions for the radial component
of acceleration for a population of planets
that varies in both mass and semi-major axis.
The distributions have been adjusted so that
they are consistent with the fraction ($\sim 10\%$)
of systems with planets with mass
between 0.3 and 10~$M_J$ and periods between 2 and 2000 days 
\citep{cumming08}.  Each system is assumed to have
at most one companion.
The central curve shows the comparison model
that has parameters listed in the first line in Table \ref{tab:tabd}.
The other curves show probabilities when a single parameter
is changed from those of the comparison model.
Lowering of the exponent associated with semi-major axis ($\beta$)
raises the probability of detecting objects
with low accelerations.  This is because the distribution
is constrained at high velocity accelerations by the
radial velocity distributions.   Lowering the exponent
of the mass distribution ($\alpha$) raises the probability of detecting
objects at all accelerations as there are more massive planets.
Truncating the planet mass distribution lowers the probability
of detecting all accelerations in the range shown here.
\label{fig:d}
}
\end{figure}

We now consider a planet mass and semi-major axis distribution consistent
with equation \ref{eqn:dP}.  We begin by normalizing the distribution
to be consistent with the fraction of systems, $\sim 10\%$, that
have planets with mass between 0.3 and 10~$M_J$ and
periods between 2 and 2000 days \citep{cumming08}.
Each systems is assumed to have at most 1 companion. 
Figure \ref{fig:d} shows the acceleration distributions using
this normalization.  These probabilities now represent the likelihood
of detecting systems with 
different radial components of acceleration.

In Figure \ref{fig:d} we show 5 probability distributions.
The comparison model is the central set of points on Figure \ref{fig:d}
with $\alpha=-1.1$, $\beta=-0.6$ \citep{marcy08}.
This model has minimum and maximum companion semi-major axes
of $a_{min}=3$~AU, $a_{max}=100$~AU.  As shown from the curve
just below the comparison model, the distribution is not
sensitive to the minimum semi-major axes.  
This is likely because the distribution is normalized
so that it extends at small semi-major axes to 
a probability consistent with the number of systems
with planets discovered with radial velocity surveys.
The comparison model has minimum and maximum planet masses
of 0.1 and 100~$M_J$.  The lowest curve shows the effect
of truncating the mass distribution to planet masses below 10~$M_J$.
The upper two curves show the effect of changing the exponents
$\alpha$ and $\beta$.  Shallower mass distributions have a larger
number of more massive objects.  
This leads to a larger number of objects predicted 
with larger accelerations.

\subsection{Mass and semi-major axis likelihoods for
a given acceleration}

\begin{figure}
\includegraphics[angle=0,width=3.8in]{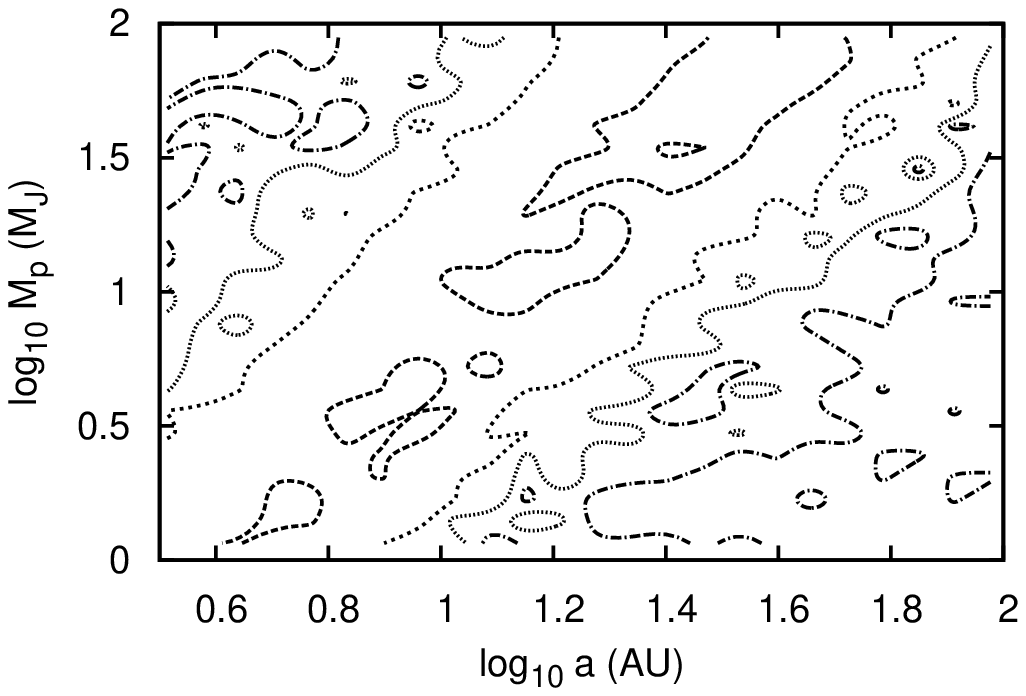}
\includegraphics[angle=0,width=3.8in]{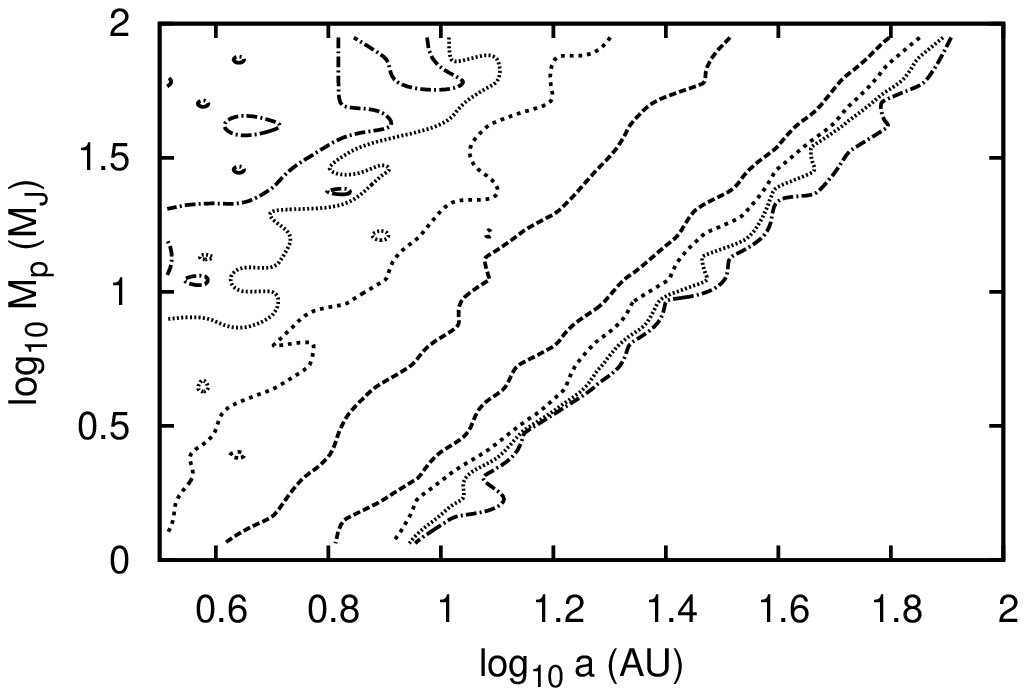}
\caption{
Probability contours showing
the most likely mass and semi-major axis of a planet
that could cause
the acceleration 4.6$\pm 0.8$ m~s$^{-1}$~yr$^{-1}$ 
observed for GL 849 \citep{butler06}. 
The contours have been estimated
using the planet and semi-major mass distribution and limits
listed in the first line of Table \ref{tab:tabd}.
Contours are shown with probabilities, $dP$ in log bins 
at 0.0003, 0.001 0.003, and 0.01.  
Flat eccentricity distributions were assumed.
a) For an eccentricity distribution with maximum
$e_{max}=0.99$.
b) For $e_{max}=0.3$.
The probability distribution is broader if the eccentricity
distribution is broad.
Objects
with masses and semi-major axes with $M_p/a^2 \approx 0.03 M_J/({\rm AU})^2$
are most likely. 
\label{fig:dm}
}
\end{figure}

For a measured acceleration  such as that
measured for GL 849 with 
$a_r = 4.6 \pm 0.8$ m~s$^{-1}$~yr$^{-1}$  \citep{butler06}
we can use our Monte Carlo code to measure the number of
systems sampled from a distribution that would be measured
with this acceleration.
The result of this using the distribution with limits
and exponents given in the first line of Table \ref{tab:tabd}
is shown with contours in a plot of $M_p$ vs $a$ in Figure \ref{fig:dm}.
Figures \ref{fig:dm}a and b show
contours for an eccentricity distribution
with $e_{max} = 0.99$, and $0.3$, respectively. 
The highest probability occurs along a line
proportional to $M_p/a^2$ or  $M_p/a^2 \approx 0.03 M_J/({\rm AU})^2$
as expected from the scaling  in equation \ref{eqn:a_r}.
\citet{wittenmyer06} illustrated upper limits
on planet mass as a function of semi-major axis 
with this scaling. Here we find that the most likely
planet mass and semi-major axis mass also have the same scaling.
The probability distribution is extremely broad implying 
that the mass and semi-major axis of the planet
cannot be tightly constrained from the probability
distribution alone.  
The distribution is less broad if
the eccentricity distribution is truncated. 
If the eccentricity distribution is broad then
10-100 $M_J$ brown dwarfs between
15-30 AU are somewhat more probable than gas giant planets.
However, HST/NICMOS observations of this object in 1997 would have detected
a brown dwarfs at a large separation, suggesting
that the acceleration is caused by a gas giant planet 
closer to the star.

\subsection{Fractions of objects
detected with specific acceleration values}

In Table \ref{tab:tabd}  we list the fraction of systems with radial component.
of their accelerations above 5, 10 and 25 m~s$^{-1}$~yr$^{-1}$. 
Table \ref{tab:tabd} shows that these fractions
are not strongly dependent on the inner semi-major axis of
the distribution, $a_{min}$ so they are not strongly dependent on 
the time span of the radial velocity survey.
The fractions are weakly dependent on the lower
planet mass $M_{p,min}$ and 
the outer semi-major axis $a_{max}$ of the distribution.
However these fractions are dependent on the outer mass
of the distribution and halve if the distribution is truncated
at 10~$M_J$ instead at 100~$M_J$.
The exponent, $\alpha$, setting the distribution dependence
on mass also affects these fractions, with lower exponents
predicting a higher fraction of stars with accelerations
above a particular value.  Lower exponents correspond
to a distribution with a larger number of massive objects.

The top part of Table 2 shows distributions normalized and consistent
with the known population of extra solar planets within 
about 5AU. 
None of the fractions of objects with accelerations
above 10 and 25 are predicted high enough to match the
fractions 23\% and 12\% estimated from
the radial velocity surveys \citep{cumming08,nidever02}.

We consider the possibility that the known distribution 
of companions based on brown dwarf companion imaging surveys
could account for the accelerations.
In the set set of lines shown in Table \ref{tab:tabd} we show
fractions of stars with accelerations 
above 5, 10 and 25 m~s$^{-1}$~yr$^{-1}$  but normalized
to a recent companion surveys in a nearby young cluster \citep{kraus08}. 
\citet{kraus08} find that 35\% of stars have companions in
the range 6--435~AU in the mass range $\sim$ 10--500~$M_J$.
We assume a stellar companion semi-major axis distribution consistent
with a flat distribution in $\log a$ or an exponent $\beta=-1$ 
\citep{heacox99,kraus08}.
For the companion mass distribution we use a shallower exponent $\alpha$
in the range $-0.5$ to 0 consistent with the
study by \citet{grether06}.
The bottom part of Table \ref{tab:tabd} shows
that fractions of objects with accelerations
above 5, 10 and 25 m~s$^{-1}$~yr$^{-1}$
predicted based on the binary companion distribution
are also not high enough to match the
observed estimated fractions of 23\% and 12\% \citep{cumming08,nidever02}.

We note that imaging surveys are necessarily incomplete, particularly
at small radii.  Consequently a higher fraction of stars 
could host  brown dwarf and stellar companions.
However the fractions shown in the third set of simulations
in Table \ref{tab:tabd} would have to be about 3 times higher
to match the estimated fractions from radial velocity surveys.

Both normalizing to the population of objects
found from radial surveys and from imaging companion surveys
predict a deficit of stellar accelerations.
To explain the fraction of objects with detected line of sight accelerations
an additional population of massive companions in the mass
range a few to 100~$M_J$ is required.
The extrapolation by \citet{cumming08} to larger
semi-major axis using the mass and semi-major
axis distribution suggested that 20\% of stars would have
gas giant planets within 20~AU. However here we find
that this extrapolation is surprisingly conservative
and probably under-predicts the number of stars with
observed radial velocity variations.

We ask how many planets would be required to predict
the estimated fraction of stars with
accelerations above 10 and 25 m~s$^{-1}$~yr$^{-1}$? 
The third sets shown in Table \ref{tab:tabd}
show the fractions $f_{10}$ and $f_{25}$ predicted
for stars assuming that 100\% host a companion
between 1~$M_J$ and 80~$M_J$ (the highest mass
that would be considered a brown dwarf).
If we restrict the companion between 4 and 20~AU 
and use size and period distribution consistent
with that estimated by \citet{marcy08} 
then $f_{10}$ and $f_{25}$ are predicted to be similar
to those estimated from radial velocity surveys.
The last line in Table \ref{tab:tabd} shows the
fractions assuming that 3 massive companions are present
in 70\% of stellar systems.    The planet inclinations in
each system are assumed to be the same in the Monte Carlo simulation.
This distribution also predicts fractions sufficiently
high to be similar
to those estimated from radial velocity surveys.
These two distribution would predict
more brown dwarfs than found by the survey by \citet{kraus08}.
Imaging surveys 
may rule out massive companions at large distances,
implying that an even larger population
of gas giant planets must exist in the outskirts of extrasolar
systems.

\section{Summary and Discussion}

In this paper we have considered the statistics of accelerations
that are detected in radial velocity surveys assuming that
they are due to massive companions with periods that exceed
the time span of the observational survey.
We find that the acceleration scales with the planet mass
divided by the square of its radius from the host star. The
acceleration is independent of the stellar mass.
As have previous studies (e.g., \citealt{patel07}), 
we conclude that detectable accelerations imply that  there
are gas giants or brown dwarfs in the outskirts  
(5 to 20~AU semi-major axis range)
of their planetary systems. 

We have run Monte Carlo simulations to predict the
statistics of the line of sight component of the radial 
stellar acceleration.
We find that
a distribution of companion masses and semi-major axes causes 
a probability function for
the line of sight component of the stellar acceleration 
in the form of a constant
minus a term proportional to the log of the acceleration.

Given an object with a measured acceleration, Monte Carlo
simulations exhibit a maximally likely product
of the planet mass divided by its semi-major axis.
The distribution is broader if the
companion eccentricity distribution is wide. 

We find that extrapolation of the planet mass and period distribution
found by radial velocity surveys within $\sim 5$ AU and companion imaging 
surveys predicts too few systems with detectable accelerations.
This suggests that a significant fraction of stars, $\sim 100\%$,  hosts 
a gas giant or brown dwarf with semi-major axis between 4 and 20AU.
Alternatively 70\% of systems could host a few gas giant planets or brown dwarfs
in a similar semi-major axis range.
A large number of planets in the outer parts 
of extrasolar systems may not be unexpected if rapid clearing
of extra solar systems is required to explain debris disks
with central clearings \citep{faber07}. 
If imaging surveys rule out distant brown dwarf companions
then we would infer that a large population of gas giants must be present
in the outskirts of extrasolar systems.

We have attempted to match an estimated fraction of stars   
$\sim$ 23\% and 12\% with line of sight 
accelerations component above 5 or 10 and 25 m~s$^{-1}$~yr$^{-1}$ 
estimated from the surveys  by  \citet{cumming08,nidever02}.
These fractions have not yet been precisely measured or reported.
Stars showing a significant fraction of a single period
or acceleration variations
(as studied by \citealt{patel07} with full orbital
analysis) could be separated
from those with constant accelerations.
We have not considered the role of jitter and uneven
time between observations and other noise sources 
in affecting the measurements of stellar accelerations.
Possible additional information could be
learned from combining astrometric and radial velocity
information for long period variations.
The study here could be refined and
broadened to more tightly
constrain the population of objects in the outer parts
of extrasolar planetary systems.
As stellar and brown dwarf companions are ruled out
from imaging surveys the fraction of objects
at low mass can be better predicted.
Better understanding of the statistics of accelerations
will probably be needed to interpret
forthcoming radial velocity and astrometric 
surveys of larger numbers of stars.


\vskip 0.2 truein
Support for this work was provided by NASA through an award 
issued by JPL/Caltech,
by NSF grants AST-0406823 \& PHY-0552695 and
and HST-AR-10972 to the Space Telescope Science Institute.
This work is based on observations made with the Spitzer Space Telescope, which is operated
by the Jet Propulsion Laboratory, California Institute of Technology under a contract with NASA.
We thank Alex Moore, Joss Bland-Hawthorne, Jian Ge and Eric Mamajek for helpful discussions.

{}

\begin{table*}
\begin{minipage}{120mm}
\caption{Fraction of single planet mass
and semi-major axis systems that
would be seen with accelerations above
5, 10 and 25 m~s$^{-1}$~yr$^{-1}$}
\label{tab:single}
\begin{tabular}{@{}lclcccc}
\hline
$M_p(M_J)$&$a$(AU)& $e_{max}$ & $f_5$  & $f_{10}$ &$f_{25}$\\
\hline
  10      &  10   &  0.0      &  0.36  &  0.22   &  0.00  \\  
  10      &  10   &  0.5      &  0.34  &  0.18   &  0.02  \\  
  10      &  10   &  0.99     &  0.29  &  0.15   &  0.03  \\  
  10      &   5   &  0.0      &  0.47  &  0.44   &  0.34  \\  
  10      &   5   &  0.5      &  0.46  &  0.43   &  0.32  \\  
  10      &   5   &  0.99     &  0.45  &  0.40   &  0.26  \\  
  5       &   5   &  0.5      &  0.42  &  0.35   &  0.14  \\  
  5       &   5   &  0.99     &  0.40  &  0.30   &  0.12  \\  
  1       &   5   &  0.5      &  0.15  &  0.02   &  0.00  \\  
  1       &   5   &  0.99     &  0.13  &  0.04   &  0.01  \\  
\hline
\end{tabular}
{\\
The fractions $f_5$, $f_{10}$ and $f_{25}$
are the fraction of stars predicted with line of sight accelerations
above 5, 10 and 25 m~s$^{-1}$~yr$^{-1}$ respectively.
These fractions were measured from Monte Carlo simulations
that considered a distribution of systems each with 1 planet
with planet mass $M_p$ (in Jupiter masses) 
(as listed in column 1), semi-major axis $a$ in AU
(as listed in column 2) and with an flat eccentricity distribution
in the range 0 to $e_{max}$ (as listed in column 3).
Orbit inclinations and mean anomalies are randomly chosen.
}
\end{minipage}
\end{table*}

\begin{table*}
\begin{minipage}{120mm}
\caption{Fraction of systems with a distribution of planet
masses and semi-major axes predicted with accelerations above
5, 10 and 25 m~s$^{-1}$~yr$^{-1}$}
\label{tab:tabd}
\begin{tabular}{@{}lccccclccc}
\hline
$\alpha$&$\beta$&$a_{min}$&$a_{max}$&$M_{p,min}$&$M_{pmax}$ &$e_{max}$&$f_5$ &$f_{10}$&$f_{25}$\\
\hline
\multicolumn{9}{l}{Normalized to radial velocity searches}\\
-1.1    &  -0.6 &   3     &  100    &   0.1     &  100      & 0.99    & 0.05 & 0.04  & 0.02   \\ 
-1.1    &  -0.6 &   3     &  250    &   0.1     &  100      & 0.99    & 0.05 & 0.04  & 0.02   \\ 
-1.1    &  -0.6 &   5     &  100    &   0.1     &  100      & 0.99    & 0.04 & 0.03  & 0.01   \\ 
-0.6    &  -0.6 &   3     &  100    &   0.1     &  100      & 0.99    & 0.19 & 0.14  & 0.09   \\ 
-1.1    &  -0.4 &   3     &  100    &   0.1     &  100      & 0.99    & 0.09 & 0.06  & 0.04   \\ 
-1.1    &  -0.6 &   3     &  100    &   0.1     &  10       & 0.99    & 0.02 & 0.01  & 0.00   \\ 
-1.1    &  -0.6 &   3     &  100    &   0.1     &  100      & 0.3     & 0.06 & 0.04  & 0.02   \\ 
\hline
\multicolumn{9}{l}{Normalized to an imaging search for companions}\\
-0.5    &  -1.0 &   3     &  250    &   0.1     &  500      & 0.99    & 0.08 & 0.07  & 0.06   \\ 
-0.5    &  -1.0 &   3     &  250    &   0.1     &  500      & 0.5     & 0.09 & 0.08  & 0.06   \\ 
 0.0    &  -1.0 &   3     &  250    &   0.1     &  500      & 0.99    & 0.07 & 0.06  & 0.05   \\ 
 0.0    &  -1.0 &   3     &  250    &   0.1     &  500      & 0.5     & 0.07 & 0.06  & 0.05   \\ 
\hline
\multicolumn{9}{l}{Assuming 100\% of stars have one massive planet or brown dwarf between 4 and 20~AU}\\
-1.1    &  -0.6 &   4     &  20     &   1       &  80       & 0.3     & 0.26 & 0.19  & 0.10   \\ 
\hline
\multicolumn{9}{l}{Assuming 70\% of stars have 3 massive planets or brown dwarfs between 4 and 30~AU}\\
-1.1    &  -0.6 &   4     &  30     &   1       &  80       & 0.3     & 0.26 & 0.21  & 0.14   \\ 
\hline
\end{tabular}
{\\
The fractions $f_5$, $f_{10}$ and $f_{25}$
are the fraction of stars predicted with line of sight accelerations
above 5, 10 and 25 m~s$^{-1}$~yr$^{-1}$ respectively.
The exponents $\alpha$ and $\beta$ correspond to a companion
distribution $dN \propto M_p^{\alpha} a^{\beta}~d M_p~da~de$.
Minimum and maximum distribution semi-major axes $a_{min}$ and $a_{max}$
are in AU.  
Minimum and maximum distribution planet or brown dwarf mass ratios
are $M_{p,min}$ and $M_{p,max}$ in Jupiter masses.  
The eccentricity distribution is assumed to be flat with
minimum and maximum $e_{min} = 0$ and $e_{max}$. \\
The top set has distributions normalized so that the number of systems
with masses between 0.3 and 10~$M_J$ and periods between 2 and 2000 days
is consistent with the 10.5\% estimated by \citet{cumming08}.
The second set has distributions normalized so that 35\% percent
have companions in the mass range $\sim 12$ to $500$~$M_J$ \citep{kraus08}.
and with semi-major axis in the range 10--430~AU \citep{kraus08}.
The exponent $\beta$ for this set is chosen from
the approximately flat distribution in $\log a$ \citep{heacox99,kraus08}.
The mass exponent $\alpha$ is chosen to be in the range
consistent with that found by \citet{grether06}.
For the top 2 sets of distributions we assume that each system has at most 1 companion.
The third set of simulations is computed assuming that all systems have
exactly one massive planet or brown dwarf with semi-major axis between 4 and 20~AU.
Exponents $\alpha$ and $\beta$ are those estimated
by \citet{marcy08}. 
For the last set we assume that that 70\% of systems have 3 giant planets or
brown dwarfs in the region between 4 and 30~AU.
}
\end{minipage}
\end{table*}

\end{document}